\documentclass[submission,copyright,creativecommons]{eptcs}
\usepackage[T1]{fontenc}
\usepackage[utf8]{inputenc}
\usepackage{verbatim}
\providecommand{\event}{F-IDE 2014}
\title{Experience in using a typed functional language for the
  development of a security application}
\author{Damien~Doligez
\institute{Inria}
\email{damien.doligez@inria.fr}
\and %
Christèle~Faure
\institute{SafeRiver}
\email{christele.faure@safe-river.com}
\and %
Thérèse~Hardin
\institute{UPMC}
\email{therese.hardin@upmc.fr}
\and %
 Manuel~Maarek
\institute{SafeRiver}
\email{manuel.maarek@safe-river.com}}
\def\titlerunning{Experience in security development}
\def\authorrunning{D. Doligez, C. Faure, T. Hardin, M. Maarek}

\begin{document}

\maketitle

\begin{abstract}
  In this paper we present our experience in developing a security
  application using a typed functional language.  We describe how the
  formal grounding of its semantic and compiler have allowed for a
  trustworthy development and have facilitated the fulfillment of the
  security specification.
\end{abstract}

\section{Introduction}

Developing an application with strong security requirements
presupposes to identify the possible threats and to address all of
them. Adverting or diverting the execution of an application is the
goal of any attack which means that its targets span almost all the
stages of development of the application and every aspect of its
execution.
On a first step in using a Formal Integrated Development Environment,
we developed an industrial security application which was
% informally
independently successfully assessed.
This development was part of a study initiated and funded by the
French Network and Information Security Agency
(ANSSI\footnote{\url{http://www.ssi.gouv.fr/}}). The overall study was
carried out by participants from industry
(SafeRiver\footnote{\url{http://www.safe-river.com/}},
Normation\footnote{\url{http://www.normation.com/}} and
Oppida\footnote{\url{http://www.oppida.fr/}}) and academia
(CEDRIC\footnote{\url{http://cedric.cnam.fr/}} and
Inria\footnote{\url{http://www.inria.fr/}}).  The subject of this
study was to determine what features of functional programming
languages can help to prevent attacks and what are the ones which
bring vulnerabilities. The results can be found on the ANSSI
website.\footnote{\url{http://www.ssi.gouv.fr/fr/anssi/publications/publications-scientifiques/autres-publications/lafosec-securite-et-langages-fonctionnels.html}}
The development part of the study was led by SafeRiver and was
informally and separately evaluated by an ANSSI team and an
independent assessor.  In this paper, we present this experience of
development. It uses a typed functional language.  We describe how the
formal grounding of the semantics of this language and its compiler
have allowed for a trustworthy development and have facilitated the
fulfillment of the security specification. We
did not address a formal description of security requirements nor
mechanically-checked proofs of their fulfillments. But, some of the
requirements have been reified onto type properties and thus
automatically checked by the typing phase of the compilation process.

\section{Requirement specification}

The project goal was to develop a secure XML validator. XML is
commonly used as a representation for data communicated by components
of a system. The parser reading the data is therefore becoming the
target of numerous attacks. There are two major sources of attacks:
the data XML file and the parser itself.  A maliciously forged XML
file given as an entry to a system can give the whole system control
to the attackers. Thus an XML validator is used to reject suspicious
XML data entries, before transmitting the XML data to any critical
parser of the system.  The specification of the XML validator we
developed is to comply to the most commonly used XSD constructions and
to do so in a secure manner, that is, to resist to direct attacks.

\subsection{Functional requirements}

The system must be able to read an XSD format and to validate
it. Then, if the XSD entry is acceptable, the tool has to generate an
XML parser according to this XSD format.  The central functional
requirement is that the produced XML parser validates a file only if
it is a well-formed XML file and if its content is valid according to
the XSD format. The application has to comply with the W3C
recommendations on XML and XSD. Most of the XSD and XML constructions
should be considered. Only those identified as ``well-known as
dangerous ones'' by the ANSSI client must be rejected.  For instance,
the wild-card XSD constructs such as \verb#<xsd:any># were discarded
as they are too permissive. We also forbid the use of numeric
character references. They allow three representations for each
standard characters (e.g., the \verb|s| character could be represented
in XML by its numeric character reference \verb|&#115;| or
\verb|&#x73;|) which could confuse an XML parser.

\subsection{Security requirements}

We made a security analysis of the XML language, the XSD language and
XML validations to define the security requirements the XML Validator
has to fulfill.

\paragraph{Availability.} A major requirement for a validation service
is not to disrupt the function it is supposed to protect. A valid
input for the service must not be rejected and must be treated in a
reasonable time. A Denial of Service attack would make use of any
input file pattern for which the validation process is
inefficient. Moreover, erroneous invalidation of input files could
lead to the whole system being moved to a degraded mode, and therefore
not carrying out its duty.

\paragraph{Integrity and confidentiality.} In addition to delivering a
trustworthy validation result, the application has to guarantee the
integrity and confidentiality of the data it manipulates. Moreover,
the application's own integrity must be guaranteed in order to serve
its security role.  A code injection attack would use any potential
flaw in the software that would permit to execute part of the input
file content.

\paragraph{Auditability.} A prevalent requirement is the ability to
demonstrate the security of the application. For the application to be
deployed it would need to be evaluated by an independent body and
deemed to comply with the appropriate standard and assurance level:
the Evaluation Assurance Level (EAL) of the Common
Criteria\footnote{\url{http://www.commoncriteriaportal.org/}} (ISO
15408) or the First level security certification for information
technologies
(CSPN\footnote{\url{http://www.ssi.gouv.fr/en/certification/first-level-security-certification-cspn/}})
in the French context.

\section{Development environment for security}

The development environment used for the development of the
application is OCaml.\footnote{\url{http://caml.inria.fr/}} The OCaml
language is strictly typed and functional. OCaml's standard
development environment comprises a type checker, a native code
compiler, and runtime system which are grounded on well studied formal
foundations. The OCaml toolkit could be extended with various tools
that are out of scope of this paper. We go into details of specific
aspects of the OCaml language and its environment which played a
crucial role in fulfilling the security requirements of the XML
validator.

\subsection{Control flow analysis enabled by the functional approach}

Analysing the control flow of an application dealing with XSD and XML
is important as these descriptive languages are complex. The
application is required to provide a negative validation result when
provided with a file that does not comply with either XML or the given
XSD as it would be a security breach. It is also required not to give
a negative validation result when provided with a valid file as it
would result in opening the possibility of Denial of Service
attack.

We designed the validation function to be purely functional. This is a
very important point. First, whatever is the occurrence of a call to
this function, its meaning is the same as it cannot depend on the
memory state. It eases a lot the traceability of the computation of
the result. Second, the formal studies of such pure typed functional
functions mathematically demonstrate that, if the function can be
typed, then its execution ends with a result of the attended type or
loops for ever. This last case being eliminated by a simple study of
recursive calls showing a decreasing measure, we can state that the
XML validator supplies a validation result.

As OCaml is not purely functional, we identified the
imperative constructions of the language and forbid their use within
the code composing the validation function. More precisely, the
following features were forbidden:
\begin{itemize}
\item The use of mutable variables as they make the control flow more
  complex to follow by proofreading,
\item The use of exceptions for nominal computation as they disrupt the
  execution flow of a program,
\item The use of non exhaustive pattern matching as it would introduce
  possible failures of the application.
\end{itemize}

To forbid such constructions, we have imposed some programming rules
to follow. These rules are of a small number. They could be verified
by proofreading of the code. Some rules could well be implemented as
static verifications on the software code or on intermediate abstract
representations provided by the compiler.

\subsection{Encapsulation as data protection}

To guarantee that neither the XML content to be validated nor the validated XSD
grammar subjacent to the used XML validator are modified by the application,
we needed to protect the data against alteration. Values are, by
default, immutable in OCaml. The memory initialisation and
manipulations are automatically performed by the compiler and the
Garbage Collector of the runtime system.

Thus there is no way of changing already acquired values  if no mutable
values are used. However, some harmful side effects are needed to
input data. We therefore decided to restrict the usage
of mutable values to isolated parts of the software and to justify
this usage in the documentation. Then, we were faced with an OCaml feature:
 the \verb#string# values are mutable in OCaml.  However, the
application makes intensive manipulations of strings and  we
needed to retain the efficiency of native \verb#string# tools. Thus,
to benefit from the native \verb#string# library capabilities
and in the meantime to forbid the mutation of these values, we have
systematically used a version of \verb#string# encapsulated in a
module. The module uses internally the OCaml \verb#string# library but does only
provide manipulation functions that cannot alter the values themselves.

To enforce this use of an alternative version of the \verb#string#, we
have followed a number of coding rules which help to identify the use
of mutable strings and which prevent the bypassing of modular
encapsulation.

Another important gain in using OCaml is the fact that data and
executable code are strictly separated in the native code produced by
the compiler. To enforce such separation, we forbid the use of dynamic
code loading and the use of the bytecode execution which are the only
situations when data could be executed. As OCaml does not have
pointers, the integrity of both data and executable code are
guaranteed. This characteristic makes impossible any injection code
attack.

\subsection{Types for traceability and prototyping}

During the validation of an XML file in accordance with an XSD
grammar, the various constructions of the XSD and XML languages are
explored. The validation process is decidable but the number of XML
and XSD constructions, their combination and the number of validation
rules specified in the W3C Recommendations increases the complexity of
both the development and evaluation of the application.
During the development of the XML validator, we heavily relied on the
union types and record types of the OCaml language, not only to encode
data but also to record the application of some W3C rules. We can say
that a large part of verifications of conformance to the W3C
recommendations was done through typing and pattern-matching.  Indeed,
the OCaml compiler systematically verifies that each operation applied
on the values of these types by pattern matching covers every possible
case, a point which improves the robustness of the application. To
reinforce this robustness throughout the development, we forbade the
use of \emph{catch-all} patterns as they create a fragile
matching. That is, in
a pattern matching for a type composed of three constructs
(e.g. \verb#type t = A | B | C#), a \emph{catch-all} pattern could be
use to discriminate every other cases but the first one (e.g.
\verb#let f = function A -> ... | _ -> ...#).  In the event of the
introduction of a forth construct later in the development, the
\emph{catch-all} pattern might  become erroneous as nothing was
known of this forth case at the time the pattern was written (e.g. if
case \verb#D# is added, the typing of \verb#f# would still hold). Not
using fragile patterns would force to make explicit the treatment of
the additional case (e.g. a function defined with explicitly listing
of all cases would have to be updated in order to type check,
\verb#let g = function A -> ... | B | C -> ...# would require an
additional case \verb#| D -> ...#).

The use of the OCaml types and the analysis performed by the type
system has proven to be decisive to systematically cover the numerous
cases of XML validation. It has also enabled the possibility of a
reliable development by prototyping stages. For each new prototype,
the types used were extended to take into account the additional
functionality.  For example, the first prototype version addressed
only the simple cases of XML patterns. The next versions added new
patterns and new rules of increasing complexity. Having just to extend
some union data types and some functions defined by pattern-matching
gave the insurance that the addition of new features was conservative
over the already done development and was easy to manage as the simple
addition of new cases.  Moreover, the OCaml type system eased the
development of new functionalities by pointing at every location in the
code that needed an update.

\section{Industrial experience and evaluation}

The development was planned in several phases. Each phase releasing a
new prototype with additional functionalities and enacting the
possibility of testing these new functionalities. The assistance
provided by OCaml type system ensured the reliability of the delivery
of each prototype.

The application was informally evaluated by an independent
expertise. The evaluation has identified a flaw (repaired in the next version) in the handling of
command line arguments which is not due to the use of OCaml. The
evaluation has not been able to identify any flaw related to its
primary purpose as a validator. Nevertheless, the fact that OCaml
applications have rarely been evaluated for security meant that there
are no tools currently available for analysing its source code and
compiled code. Commonly used binary code analysers are of little help
for the auditor. This lacking might slow a formal evaluation.

\section{Conclusion}

The experience of developing an XML Validator with OCaml has
demonstrated that it is a suitable environment for security
developments. Moreover, the ability to monitor the development of the
application with prototyping phases is of high industrial
relevance. The formal grounding of the OCaml language and type system
makes OCaml and its compiler a high level development environment. The
tooling available is nevertheless limited, especially in an industrial
context.  Thus, it would be interesting to develop an IDE for OCaml,
dedicated to critical software production. Some recommendations on
this usage of OCaml are given in the study LAFOSEC, which precises
what are the features of Ocaml which must be kept and what are the
ones that must or can be dropped, what extensions and new libraries
are needed. To reach the possibilities of a F-IDE, a logical component
should be integrated to such an IDE. Focalize can be a very good model
of this integration as it is based on the functional kernel of
Ocaml. Its current compiler ensures a complete computational and
logical treatment of the source development, which issues Ocaml and
Coq codes, very close to each other and to the source code. It
guarantees also a very large sharing of object codes and it provides
automatic documentation and some testing possibilities. To summarize,
developing this XML validator within a F-IDE \emph{à la} Focalize
should end with a product able to pass all the controls needed by an
evaluation at the upper-rate level of security standards.

% ANSSI
\nocite{LaFoSec_langages-2011,LaFoSec_securite-2011,LaFoSec_execution-2011,LaFoSec_outils-2011,LaFoSec_recommandations-2011,LaFoSec_xsvgen-2012}
% XML XSD
\nocite{W3C_XML_1.1-2006-aug,W3C_XSD_part1-2011-jul,W3C_XSD_part2-2011-jul}
% OCaml
\nocite{OCaml3.12-2010}
% FoCaLize
\nocite{CESAR08}
\bibliographystyle{eptcs}
\bibliography{f-ide-14}

\end{document}